# Critical state analysis in $MgB_2$ bulk by means of quantitative MO technique

L. Gozzelino, F. Laviano, D. Botta, A. Chiodoni, R. Gerbaldo, G. Ghigo and E. Mezzetti

*INFM - U.d.R Torino-Politecnico; INFN - Sez. Torino*
*Politecnico di Torino, c.so Duca degli Abruzzi 24, 10129 Torino, Italy*

G. Giunchi, S. Ceresara, G.Ripamonti, M. Poyer

*Edison S.p.A.*
*Foro Buonaparte 31, 20121 Milano, Italy*

**Abstract**

An extended magneto-optical (MO) analysis of samples cut from high-density pellets of $MgB_2$ is reported. For sake of comparison some magnetic measurements and critical values are also shown. Notwithstanding the fact that the optical and SEM images exhibit grains with different shape and size, all the samples investigated by MO analysis (three specimens of different shape and size) enter as a whole into a critical state. The current at different temperatures and fields, in a phase zone accessible to MO are calculated by means of a quantitative analysis. The analysis, based on the inversion of the Biot-Savart law, corresponds to three regimes: under-critical, critical and over-critical, respectively. A possible qualitative interpretation of the absence of magnetic granularity is given in the framework of a critical state likely reached by a network of strongly coupled Josephson Junctions.

**Introduction**

The magnesium-diboride is an astonishing strong intermetallic superconductor. Large current carrying capability and strong magnetic response were expected for this compound since its recent discovery [1,2]. High density $MgB_2$ bulk pellets have been prepared starting from the elemental compounds B (99.5% of purity) and Mg (99.9% of purity), after their reaction in a sealed stainless steel container, lined with a Nb foil. The thermal treatment was performed for two hours in the range of 850°C-950°C. The preparation procedures of highly densified $MgB_2$ pellets are reported elsewhere [3].

In this paper we study, by means of a quantitative magneto-optical technique (MO), the magnetic field distributions and the macroscopic supercurrent flow across samples cut from the pellets. Beside the MO analysis [3,4], performed inside a modified commercial cryostat whose temperature is controlled from 4 to 300 K, a magnetic characterization was also made by means of a commercial extraction magnetometer. From the full magnetic analysis some basic information about the sample quality is extracted and reported here.

For the present investigation four samples were cut from a polycrystalline pellet. The sample #1 was shaped as a disk and used for susceptibility and magnetic hysteresis characterizations. The sample #2 has irregular parallelepiped shape and measures (10x10x2) $mm^3$. Although it is too large to allow a quantitative MO analysis, it was selected to show how the general aspects related to the flux penetration do not depend of the sample dimension. A third near-cubic sample of dimension suitable to be observed with a single intensity/field calibration procedure, was selected and cut from the pellets. Its dimensions are (1.4x1x0.8) $mm^3$. The sample #4 size is (1x3x1.9) $mm^3$ and the shape is a quasi-regular parallelepiped. This last specimen was chosen for the extended analysis in order to show the following flux penetration characteristics:

i) into a continuous -bulk granular structure

ii) inside a macroscopic crack

iii) into a zone where the streamline is perturbed by the channelling of the flux into a “bottle neck” [13].

## Results

### 1) *Magnetic characterisations.*

In Fig.1 the temperature dependence of the a.c. susceptibility is shown. In the inset the first derivative of the real part of the a.c. susceptibility is plotted in order to gain a better evaluation of the critical temperatures and transition width, reported in the caption.

In Fig.2 magnetisation hysteresis loops measured at different temperatures are plotted. In the inset the magnetisation cycle at 36 K is shown. The penetration field, $\mu_0H$, at 36 K as well as the critical current density, Jc, at the same field and temperature are also evaluated for sake of comparison with the MO analysis. It results $\mu_0H=0.12T$, Jc $7.3\cdot10^7$ A/m$^2$. All data have been normalized to the whole volume of the sample and have been corrected to account for the demagnetisation effects.

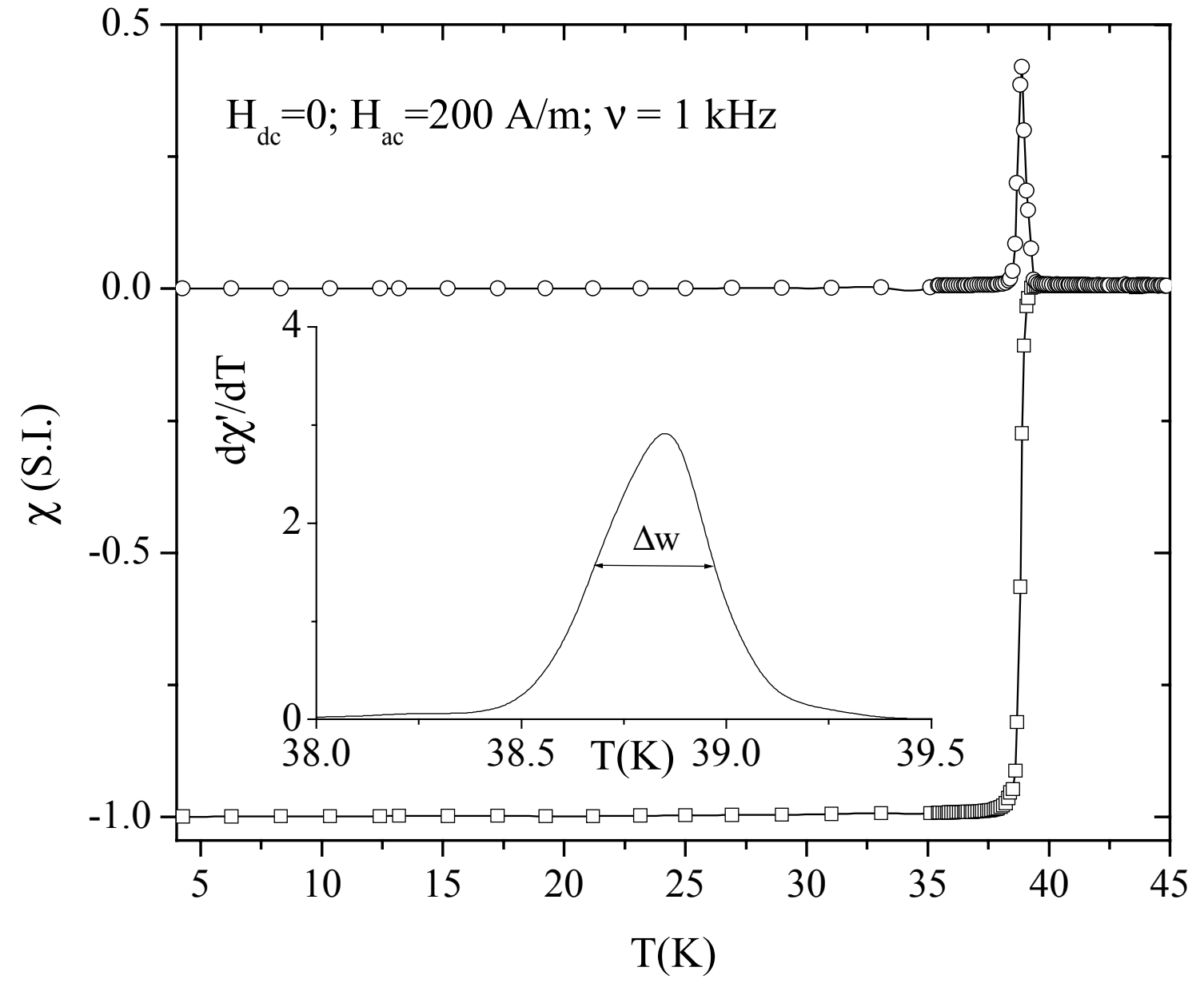

**Figure 1**. Temperature dependence of the a.c. susceptibility. Inset: the first derivative of the real part of the ac susceptibility. The critical temperature $T_c$ evaluated as the temperature corresponding to the maximum, is 38.9 K . The onset temperature $T_0$ is 39.4 K. The FWHM, Δw, is 0.3 K.

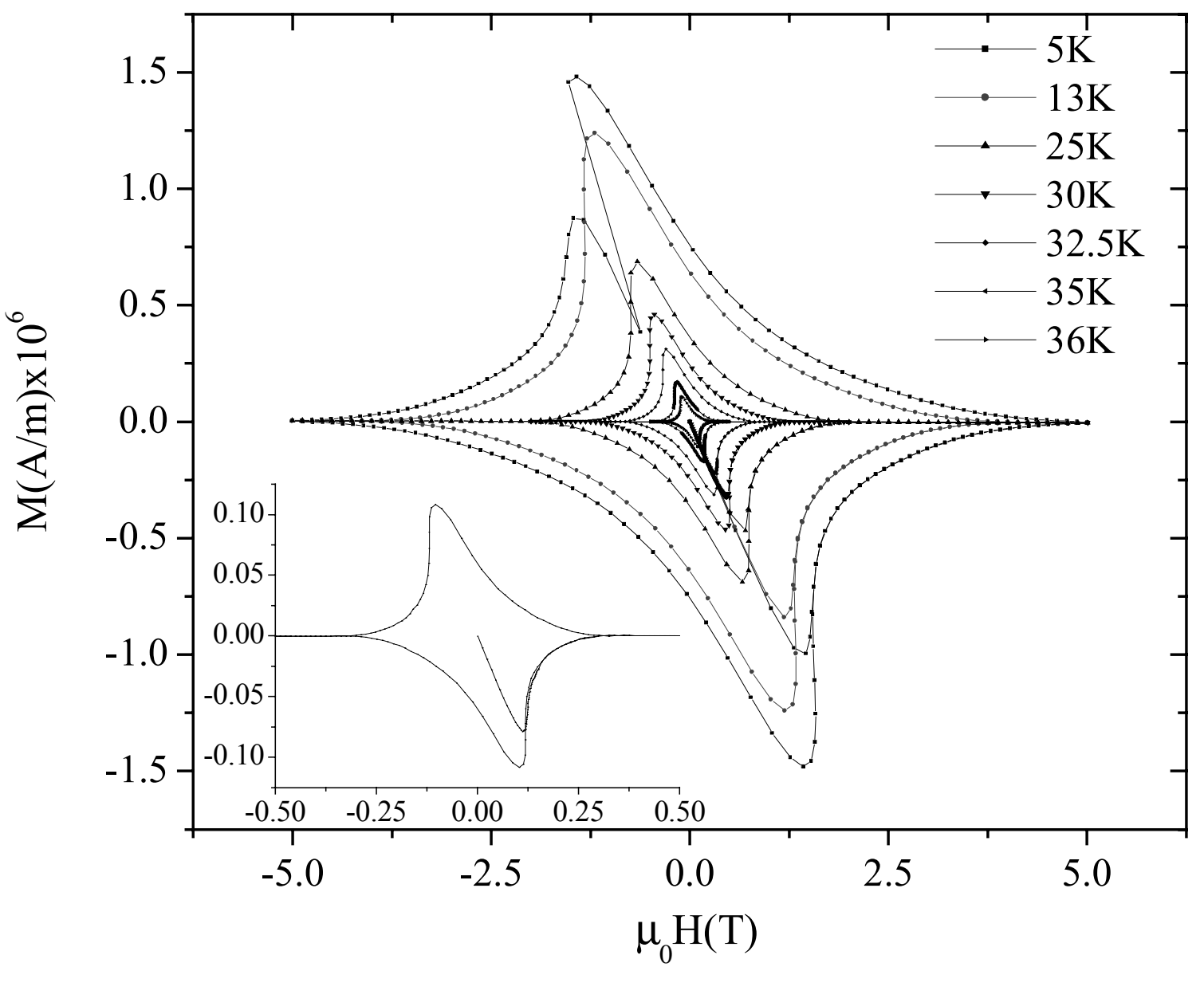

**Figure 2**. Magnetization hysteresis loops measured at 5, 13, 25, 30, 32.5, 35, 36 K (from outer to inner). All the curves here show a symmetric behaviour indicating the importance of bulk current instead of surface shielding current. In inset the hysteresis cycle at 36 K is shown, not distinguishable in the main frame. The penetration field at 36K is 0.12 T, the critical current density at the same field and temperature is $7.3\cdot10^7$ A/m$^2$.

2) ***Magneto optical analysis***

Here we report on two kinds of measurements

- zero field cooling, ZFC (the sample is cooled up to 4 K in zero field, starting from a temperature above the transition, then warmed up and measured at different temperatures and field)
- remnant state, RS (the sample is cooled in zero field, after that a given field is applied and removed; the image is taken as soon the field is removed).

Figure 3a represents the MO picture of sample #2 at 35K in the remnant state after switching off an external magnetic field of 0.135 T. As outlined before the sample size is too large to observe the entire geometry, then it was necessary to merge several pictures.

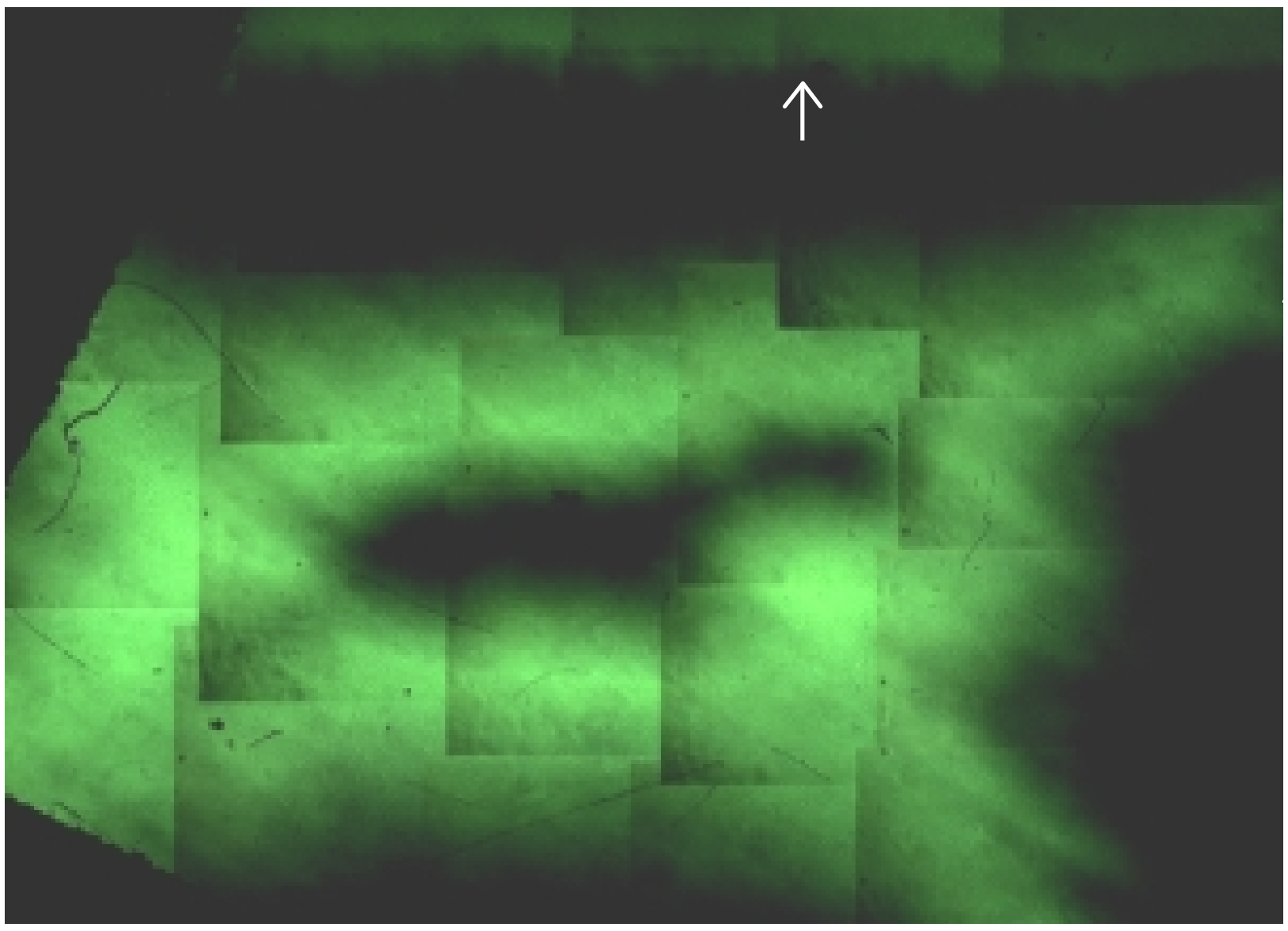

**Figure 3a**. Light intensity picture (RS) of the sample #2 at 35 K after an applied field $\mu_0 H$ = 0.135 T. An edge of the sample is visible in the upper part (arrow); two edges of the indicator film are visible on the left. In the centre there is a totally screened region showing the main characteristics of a critical state. Minor deviations from ideal critical state penetration are due to the irregular shape of the sample.

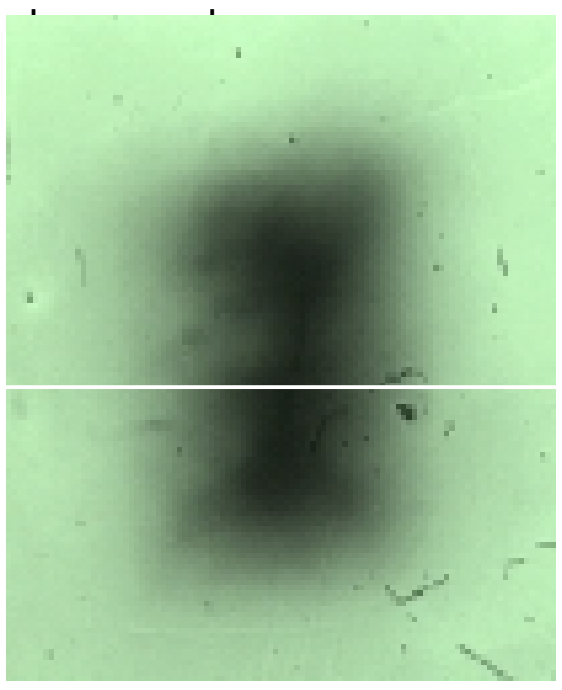

**Figure 3b**. Light intensity picture (ZFC) of the sample #3 at 35 K with an applied field $\mu_0 H$ = 0.075 T.

**Figure 3c**. Induction intensity profile along the straight line in Fig. 3b.

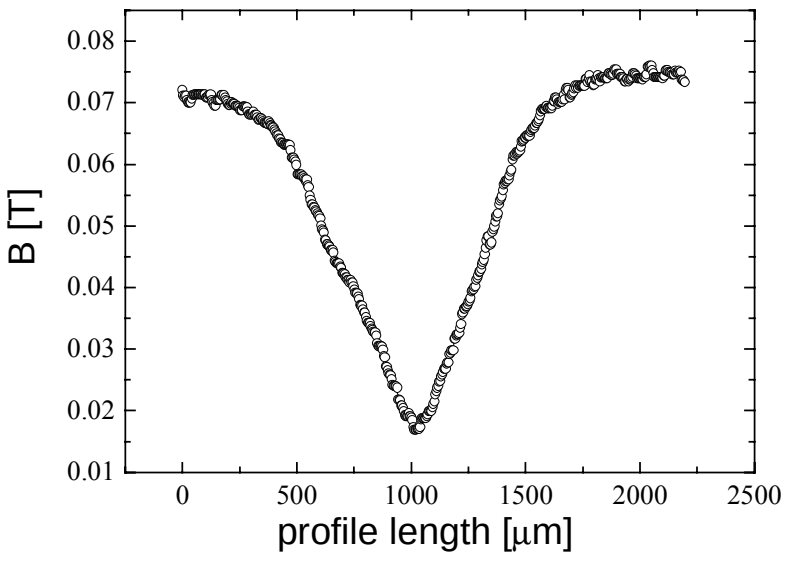

*In spite of its granular nature and relatively high temperature, the sample behaves as a whole.*
The field penetrates from the edge reaching the position of maximum light intensity. The flux front is geometrically well connected and presents discontinuity lines, above considered, corresponding to the sample corners. This pattern is characteristic of a “crystal” in a critical state regime, remnant state [6-7]. In early work [2] this feature was not directly observed. In sample #3 (Fig 3b) the critical state characteristic pattern, taken in zero field cooling mode, mirrors its nearly perfect regular shape, confirmed by the field profile in correspondence to the straight line (Fig. 3c).
Henceforth a more extended analysis is reported concerning sample #4. The granular microstructure and a macroscopic crack are clearly visible in the optical figure (Fig 4a) as well as in the SEM image, Fig 4b. The macro-defect allows studying the interplay of sample imperfections on different scales.
The MO picture in Fig 4c, taken at 11 K with 0.12 T, demonstrates that at low temperature and low field the sample is fully screened with the exception of the crack position where the field penetrates like a not superconducting region. The left side part is almost separated from the rest of the sample, has a regular trapezoid form and play a major role in the present investigation and conclusions. To compare MO measures with quantitative knowledge about the critical state behaviour we focused the analysis on this part of the sample at temperatures large enough to observe an appreciable flux penetration.

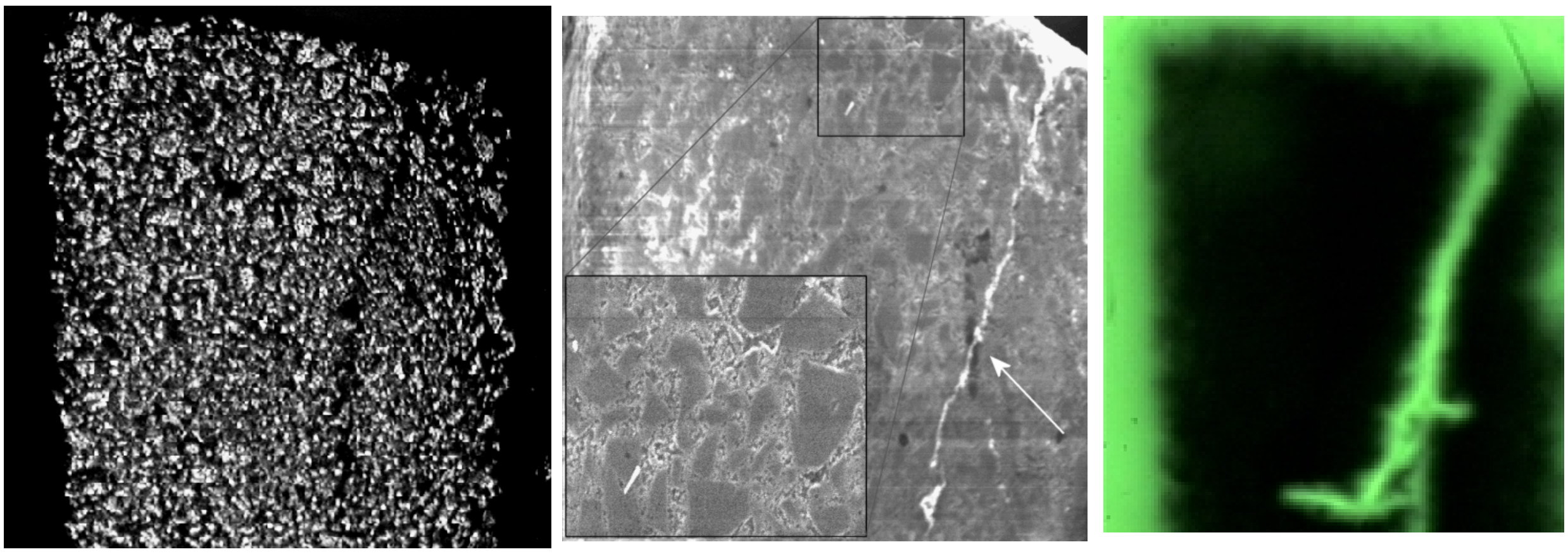

**Figure 4.** a) Optical image of the upper part of the third crystal. b) Scanning electron microscopy image (SEM magnification 50X). The arrow indicates the crack. In the insert a magnification of the square circumscribed part of the sample (100X). c) MO light intensity image taken at 11 K ( zero field cooling mode) with an applied field $\mu_0$H=0.12 T.

Fig.5 shows a MO image taken in zero field cooling mode. The light intensity (a) and the field induction value along perpendicular direction to the indicator film (b) are reported. The image is taken at 35K with an applied field of 0.051 T. The magnetic flux penetrates the sample from the edges (middle part). Due to the fact that the crack acts like a normal region thus it is considered as an edge of the trapezoid part.
The magnetic image shows the typical horn-shaped pattern which correspond to the discontinuity lines**,** first observed by Schuster et al. [6] **(**d+ lines) and attributed to the current path bending. According to the calculation of Brandt [7] this flux profile pattern is due to a critical state of nearly constant critical current density $J_c$.
Further investigation on current flow patterns is made by using the inverted 2D Biot-Savart model [8-9].

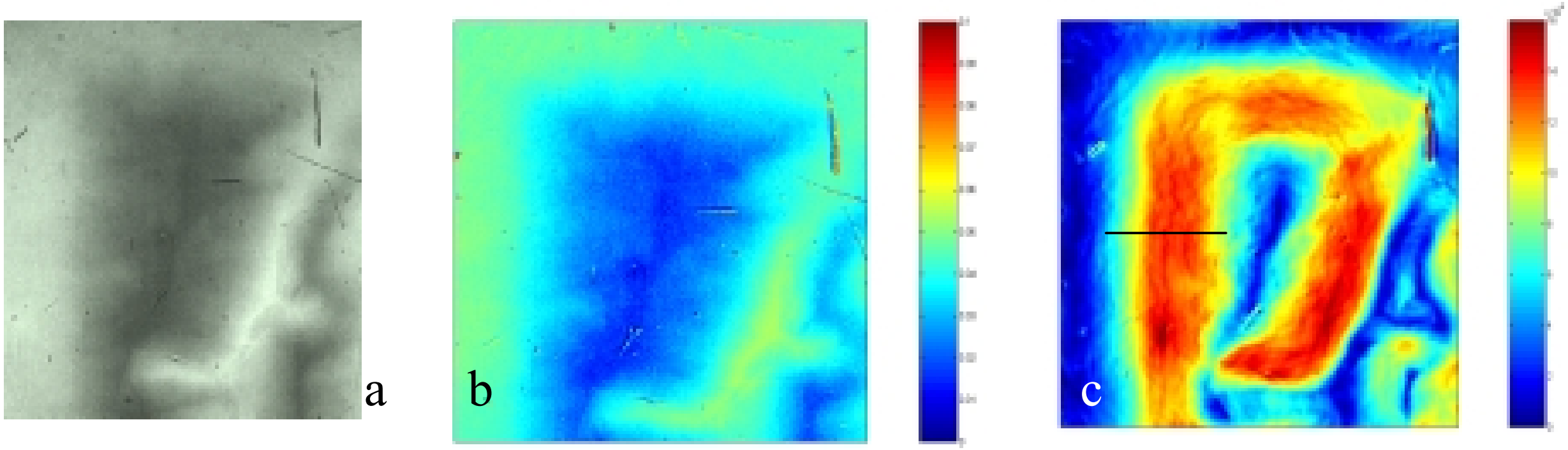

**Figure 5**. a) Light intensity map taken in zero field cooling regime at T=35 K and B=0.051 T. b) Field map. The right bar indicates the field scale. c) Local current intensity map. The line indicates the position where the current values were been taken. The bars indicate the field and current scales, respectively [T] and [Am$^{-2}$]. For pattern shape comparison see Ref. 13, Fig 6j.

Figure 5 shows a current intensity map relative the previous measure. The current flows in the regions penetrated by the magnetic field. At the corners the current exhibits sharp bending and depletion of current-streamlines density is in qualitative accordance with the calculations of Ref. 4, Fig.5, concerning the critical state. Inside the crack, the induced field is the sum of the two contributions coming from the supercurrent flowing in both the crack banks because, as expected, the crack is a not superconducting macro-channel. Then the shielding currents in the boundary region are larger than in the other side. Moreover, in the remnant state , shown above in Fig.7, the flux is totally absent in the canyon region in the full range of temperatures, confirming the previous argument. In the bottom left part of the sample, the constriction creates a superconducting bridge where both the flux gradient (Fig 4b) and the supercurrent are large. A weaker superconducting bridge is observable in the right part near the crack. The ratio between the supercurrent flowing in the two bridges (right divided left current value) is 54.5%.
The trend of the local current versus $\mu_0 H$ and T has been evaluated from the images in three regimes: below the critical state, in the critical state (when reached) and above. The quantitative results are shown in Fig 6. The critical state corresponds to the maximum of the curves. In the inset of fig 5 the zone under quantitative evaluation is shown.

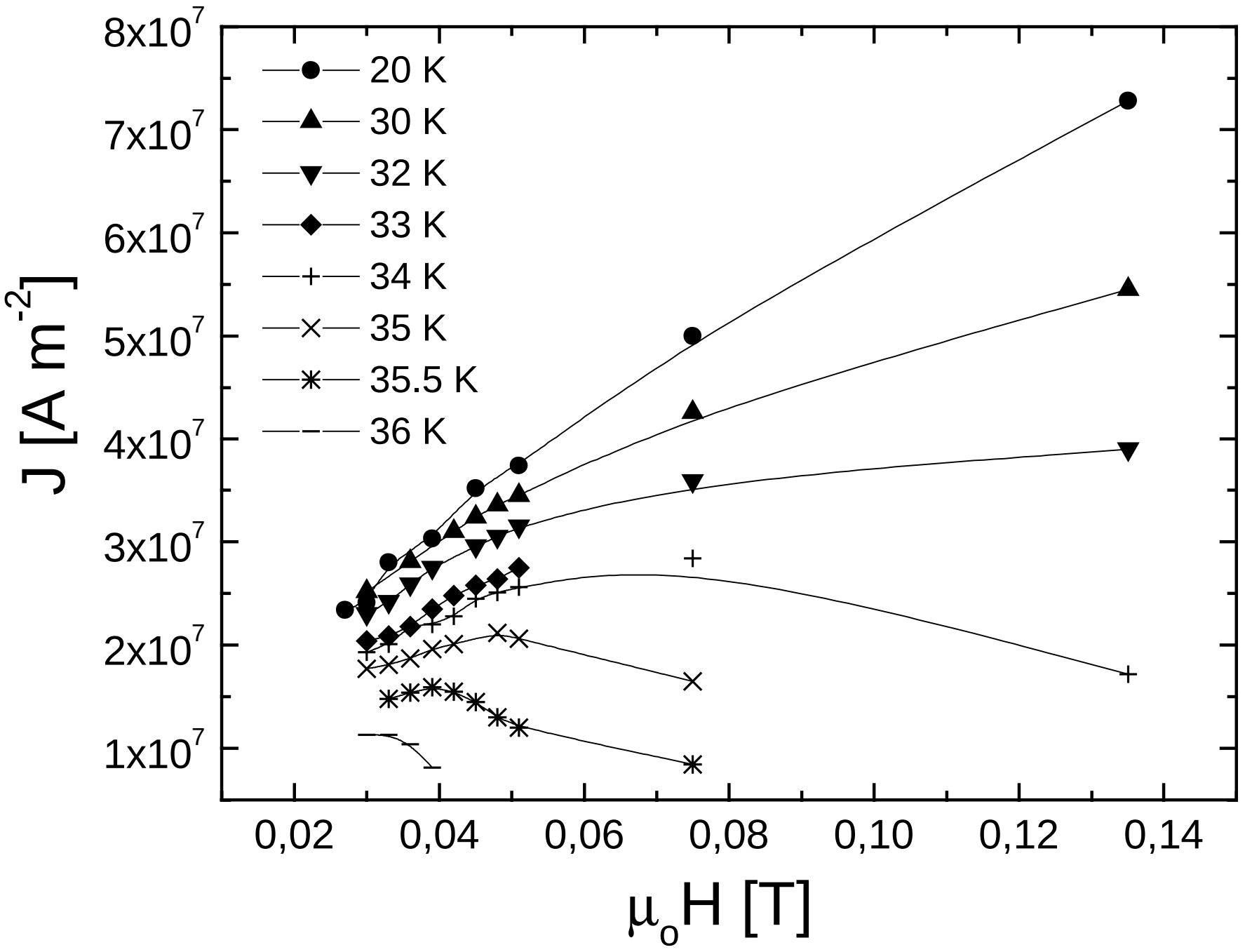

**Figure 6**. Local current as a function of the local magnetic field at different temperatures in sample #4. The values are calculated by means of the inverted 2D Biot-Savart model.

In a framework of comparison between these results and those obtained by magnetic analysis, (Fig. 2 inset and Fig. 2 caption), it must be pointed out that the magnetic analysis is a volume analysis while the MO analysis is mainly sensitive to the upper surface-current distribution. Thus if we take into account the experimental errors (≤15%) this coarse agreement is a further argument in favour of the strongest intercoupling of the material.
At low temperature (below 33 K), the fields observable by MO are not large enougth to cause the critical state. However between 34 K and $T_c$ the critical state is clearly present. When it is reached, the supercurrent exhibits a maximum. This value then decreases with increasing field. For type II granular superconductors it is expected that fluxons enter as Josephson vortices trough grain boundaries and this occurs at field lower than the Hc1 relative to the grains; Chen et al.,[11,12], demonstrate that, above a current threshold, a transition from the vortex state to the critical state sets up inside a network of Josephson Junctions (JJs). In that case, the superconductor behaves as a whole. Thus our observations of granularity from one hand, and strong coupling from the other hand, are consistent if we assume that the microstructural grain-boundary network has a magnetic counterpart in a disordered JJ network , thought our resolution (~4 μm) does not resolve single vortices. The network would be so framed that the grain boundaries are both flux-flow channels and pinning centers. Namely the measure in remnant state (Fig. 7) is in full qualitative accordance with results in Ref. 11 concerning the remnant state of the JJs array in critical conditions.

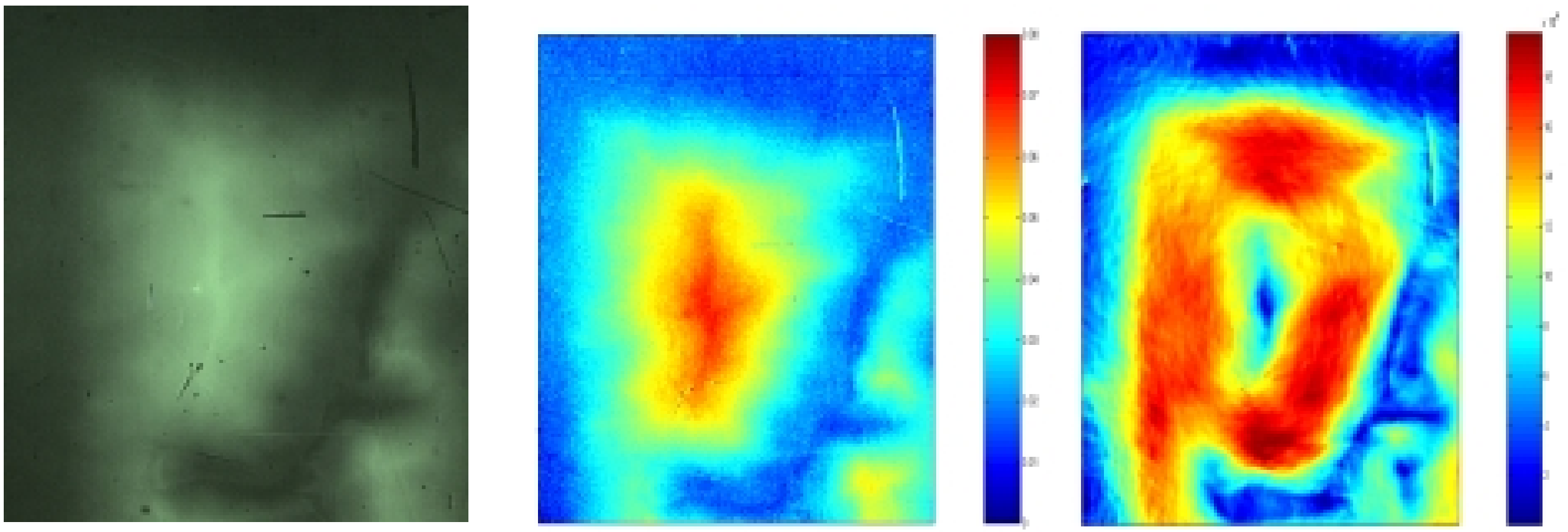

**Figure 7.** Remnant state images of sample three at 35 K and maximum applied field of 0.135 T. a) Light intensity. b) Magnetic flux distribution pattern. The right bar indicates the field scale [T]. c) Local current intensity map. The bar indicates the current scale [$Am^{-2}$].

## Conclusions

The results of a MO analysis of good quality $MgB_2$ bulk-samples, together with some basic magnetic characterizations have been reported.

In the measurements we observe for the first time**,** to our best knowledge, in $MgB_2$, a critical state. All the observed samples, however large, enter in this state as a whole. We chose a particularly feature-rich sample for the quantitative analysis. The local variation of the supercurrent magnitude with field and temperature in under-critical, critical and over-critical state in the point-phase accessible to the MO analysis was calculated. With particular reference to remnant-state measurement (Fig. 7), we argue that sample granularity from one hand, and strong pinning / high percolative currents from the other, can become consistent in the framework of the Brandt-Chen picture of a strong coupled JJ network. A particular coupling strength of the network would determine the macroscopic current magnitude and the flow patterns.

## Acknowledgement

The authors wish to thank E. Bennici for SEM image.

## References

[1] J. Nagamatsu, Nature 410 (2001), 63.
[2] D.C. Larbalestier et al., Nature 410 (2001),187.
[3] Edison S.p.a., patent pending.
[4] A. Polyanskii et al., Proc. of NATO Advanced Research Workshop (Sozopol, Bulgaria, Sept. 1998) ed. by I. Nedkov and M. Ausloos, NATO Sciences Series 3: High Technology, Vol. 72 (1999) p. 353.
[5] E. Mezzetti et al., Physica C in press.
[6] Th. Schuster et al., Phys. Rev. B **52** N°21 (1995), 15621.
[7] Th. Schuster et al., Phys. Rev. B **52** N°14 (1995), 10375.
[8] B.J. Roth, J.Appl.Phys . **65** (1989), 361.
[9] Ch. Jooss et al., Physica C **299** (1998), 215.
[10] D. Yu. Vodolazov, I.L. Maksimov, E.H.Brandt, cond-mat/0101074.
[11] D. -X. Chen et al., Phys. Rev. B **53** (1996), 6579.
[12] D. -X. Chen et al., Phys. Rev. B **56** (1997), 2364.
[13] A. Forkl et al., Phys. Rev. B **52** (1995), 16130.